\documentclass[twocolumn,preprintnumbers,amsmath,amssymb,superscriptaddress,prb]{revtex4}
\usepackage{epsfig,amsopn}
\usepackage{graphicx}
\usepackage{epstopdf}
\usepackage{sidecap}
\usepackage{color}
\usepackage{amsmath,amssymb}
\usepackage{amsthm}
\usepackage{enumerate}
\newcommand\mb{\mathbf}

\begin{document}
\title{Brillouin-Wigner Theory for Floquet Topological Phase Transitions in Spin-orbit Coupled Materials}
\author{Priyanka Mohan}
\affiliation{Harish-Chandra Research Institute, Chhatnag Road, Jhunsi, Allahabad 211 019, India.}
\affiliation{Homi Bhabha National Institute, Training School Complex, Anushaktinagar, Mumbai, Maharastra 400085, India.}
\author{Ruchi Saxena}
\affiliation{Harish-Chandra Research Institute, Chhatnag Road, Jhunsi, Allahabad 211 019, India.}
\affiliation{Homi Bhabha National Institute, Training School Complex, Anushaktinagar, Mumbai, Maharastra 400085, India.}
\author{Arijit Kundu}
\affiliation{Physics Department, Technion, 320003, Haifa, Israel}
\author{Sumathi Rao}
\affiliation{Harish-Chandra Research Institute, Chhatnag Road, Jhunsi, Allahabad 211 019, India.}
\affiliation{Homi Bhabha National Institute, Training School Complex, Anushaktinagar, Mumbai, Maharastra 400085, India.}

\begin{abstract}
We develop the high frequency expansion based on the  Brillouin-Wigner (B-W) perturbation theory for driven systems with spin-orbit coupling which is applicable  to the cases of silicene, germanene and stanene. We compute the effective Hamiltonian in the zero photon subspace not only to order $O(\omega^{-1})$, but by keeping all the important terms to order $O(\omega^{-2})$, and obtain the photo-assisted correction terms to both the hopping and the spin-orbit terms, as well as new longer ranged hopping terms. We then  use the effective static Hamiltonian to  compute the phase diagram in the high frequency limit and compare it with the results of direct numerical computation of the Chern numbers of the Floquet bands, and show that at sufficiently large frequencies, the B-W theory high frequency expansion works well even in the presence of spin-orbit coupling terms.
\end{abstract} 
\pacs{}
\maketitle

\section{Introduction}  
Topological insulators and topological phase transitions~\cite{topologicalreviews} have been in the forefront of research in the last several years. More recently, it has been realized that driving systems periodically is an effective way to obtain and control topological phases~\cite{Oka_2009, Kitagawa_2010,Lindner_2011,Dora_2012}. In the last few years, the concept of engineering such periodically driven systems, often called Floquet systems, has gained prominence, particularly due to the feasibility of experiments in solid state\cite{Wang_2013} as well as in photonic\cite{Rechtsman_2013} and cold atom systems\cite{Jotzu_2014}. Floquet topological systems have been studied extensively to predict non-equilibrium Majorana modes\cite{Jiang_2011,Kundu_2013,Li_2014}, non-trivial transport properties\cite{Gu_2011,Kundu_2014,Titum_2015,Farrell_2015} as well as to control the band-structure\cite{Lindner_2011,Kundu_2016,Klinovaja_2016}. 

Despite this  progress, there remain many unresolved questions involving driven topological systems, mainly because the presence of the driving implies that the system is out of equilibrium. With the lack of energy conservation, the bands in a driven system can be characterized by \textit{quasienergies}~\cite{Sambe_1973}. But the standard picture of assuming that the quasi-energy levels are similar to the usual energy levels of a band is not quite right  because the distribution function for the electrons in the quasi-energy bands cannot be assumed to be the usual Fermi distribution function. Furthermore, a driven system has a much richer topological phase structure  than its static counterpart\cite{Rudner_2013} and may even possess phases that have no analogue in the static system~\cite{Titum_2016}. This has led to the proposal of characterising the topological indices  of a periodically driven topological insulator as   a combination of winding numbers  instead of a single Chern number.

 \begin{figure}
 \begin{center}
  \includegraphics[width=0.4\textwidth]{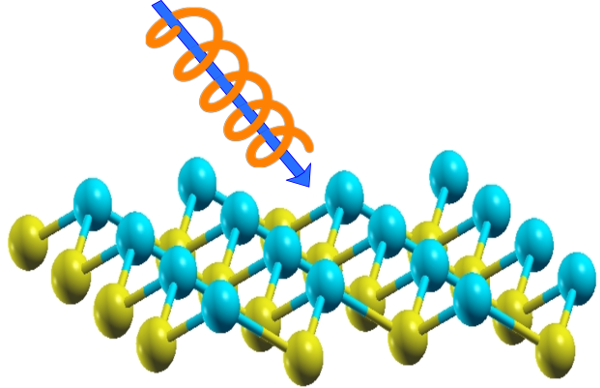}
   \end{center} 
   \caption{A class of materials, such as silicene, stanene and germanene has 2D electrons in a buckled structure. Sites on a blue (upper) layer and a  yellow (bottom) layer together form a lattice.} 
   \label{setup}
 \end{figure}

Motivated by graphene, much of the early work~\cite{Kitagawa_2011,Gu_2011,Kundu_2014,Wang_2016,Usaj_2014} on Floquet topological insulators has been on understanding the topological features of periodically driven  tight-binding models on a honeycomb lattice. However, it is also of interest to extend the work to include spin-orbit coupling terms and buckling terms which are of relevance to materials like silicene, germanene and stanene~\cite{Drummond_2012,Liu_2011b,Liu_2011a,Ezawa_2011a,Ezawa_2011b}. Although these materials are intrinsic topological insulators and their band gap can be tuned by an external gate voltage, fairly large electric fields are required to tune the materials between topological and normal insulators. The enhanced tunability offered by using light as a driving force may allow us to access many more topologically distinct phases in these materials. 

A simple theoretical idea that has been used in the field of Floquet systems is to realize that at very high frequencies, when the frequency of the drive is larger than the band-width, the system cannot follow the rapid oscillations of the external drive and hence, the effective Hamiltonian is just the time-averaged one. An effective Hamiltonian is then  systematically constructed  using perturbation theory,  at high frequencies, to include virtual photon absorption and emission processes to give corrections of $O(\omega^{-n})$, where $\omega$ is the frequency. Here, it has been shown that at least at high frequencies, in models like graphene, the assumption that the quasi-energies can be treated as usual energy levels works well.

In an earlier work, Ezawa\cite{Ezawa_2011a} investigated  photo induced phase transitions in silicene and showed that at high frequencies,  various new phases such as the quantum Hall insulator, spin-polarized quantum Hall insulator, spin polarized metal and spin-valley polarized metal are realized. However, his study was restricted to high frequencies of $O(\omega^{-1})$ in the high frequency expansion and also to low energies, close to the Dirac cone.  In this paper, we study a systematic Brillouin-Wigner expansion\cite{Mikami_2016} of the effective Hamiltonian of systems with a spin-orbit coupling term, and obtain the effective Hamiltonian to $O(\omega^{-2})$, 
without restricting ourselves to the low energy limit.  To obtain  the phase diagram, which should be qualitatively applicable to  all materials with spin-orbit couplings such as 
silicene, germanene and stanene,  we keep the spin-orbit term small but arbitrary.  Thus we are able
to access many more phases in the spin-polarized, buckled systems.

To be more specific,  the plan of our paper is as follows. Since our aim is to extend the B-W theory for high frequency expansion to materials which also have spin-orbit coupling,
we start, in the next section, with a brief review of the expansion procedure  which will also serve to define our conventions. Then, in section ~\ref{sec:III}, we write down  our model of the materials of interest,  which is a tight-binding model on a honeycomb lattice with next nearest neighbor spin-orbit terms. Since many of the materials of interest have a buckled structure, we
also include a staggered sub-lattice potential term. In momentum space, this is a 4-band model with real spin as well as pseudo-spin or valley indices. In section ~\ref{sec:IV},  we proceed to generate the effective Hamiltonian in the projected  zero-photon subspace of the Floquet Hamiltonian by using the B-W perturbation theory order by order in $1/\omega$.  We show that at each order, the effective Hamiltonian  has longer and longer ranged hoppings. For high frequencies, we truncate our expansion to $O(\omega^{-2})$. The low-energy limit of the effective Hamiltonian near the Dirac points in the Brilloiun zone is discussed in section ~\ref{sec:V}. 
In Sec~\ref{sec:VI},  we show numerical evidence of topological phase transitions in the effective Hamiltonian as a function of the amplitude and frequency of the driving force,  as well as a function of the staggered potential 
and compare it with  exact results. We conclude with a discussion of where we expect the B-W expansion to give a reasonable  approximation
 of the time-dependent Hamiltonian, - $i.e.$, we obtain a range of validity for the parameters of the theory, where we can expect the B-W expansion to 
provide a reliable time-independent Hamiltonian.

\section{B-W high frequency expansion}\label{sec:II}
The B-W perturbation theory has been described in Ref.~\onlinecite{Mikami_2016}  to obtain the high frequency effective Hamiltonian for  periodically driven systems. In comparison with other similar high frequency expansions, like Floquet-Magnus~\cite{Casas_2001,Mananga_2011} and van Vleck~\cite{Bukov_2015,Eckardt_2015} perturbation theory, the B-W expansion has far fewer terms at higher orders. Moreover, the B-W theory has a simple recursive technique to compute higher order terms which is often less cumbersome than the other expansions.
 
In this paper, we will only use the B-W theory, since with the addition of spin-orbit couplings, we have  even more terms and  the recursive technique can be conveniently used to compute the higher order terms. We start with a time periodic Hamiltonian $H(\tau + T) = H(\tau)$, where $T=2\pi/\omega$ is the period, given by its Fourier components
 \begin{align}
 H_{n} = \int_{0}^{T}\frac{d\tau}{T} H(\tau) e^{in \omega\tau}. \label{hf}
 \end{align}
The B-W perturbation theory can now be used to obtain the effective Hamiltonian order by order in
$1/\omega$ as\cite{Mikami_2016}
\begin{align} 
  H_{\rm{BW}}=&\sum^{\infty}_{n=0} H^{(n)}_{\rm{BW}} \quad
\end{align}  
where, the first few orders are:
\begin{align}\label{BWexp}
  H^{(0)}_{\rm{BW}}=& H_{0} \nonumber  \\
  H^{(1)}_{\rm{BW}}=&\sum_{n\ne0} \frac{H_{-n}H_{n}}{n \omega}  \nonumber \\
 H^{(2)}_{\rm{BW}}=& \sum_{n,m\ne0} \left( \frac{H_{-n}H_{n-m} H_{m}}{n m \omega^2}-\frac{H_{-n} H_{n} H_{0}} {n^2 \omega^2} \right).
\end{align}
 
 \begin{figure}
 \begin{center}
  \includegraphics[width=0.35\textwidth]{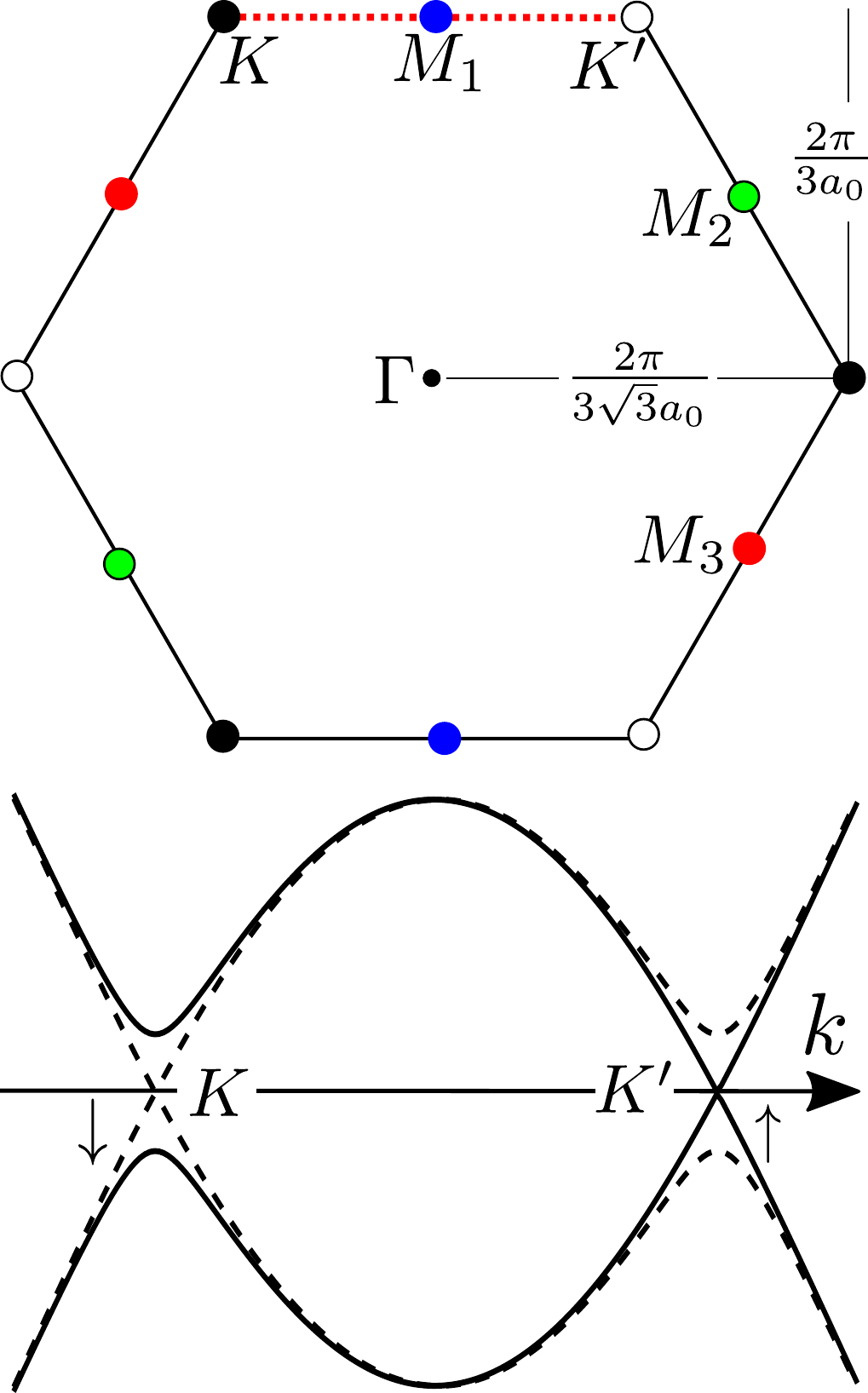}
   \end{center} 
   \caption{(a) The hexagonal Brilloiun zone for the model that we consider with various high-symmetry points (discussed in the text) marked.  (b) The band structure with momenta along the red dotted line in (a)  in the presence of spin-orbit coupling and a staggered electric field $E_z$. 
 It is possible to have Dirac nodes for different spins at different valleys. A spin-orbit coupling constant of $\lambda=0.1t$ and a  staggered potential of $lE_z=0.1 t$ have been used to obtain the schematic diagram presented here.} 
   \label{static}
 \end{figure}
 
As an example-system, it is useful to consider electrons in a honeycomb (hc) lattice (say, graphene) irradiated by circularly polarized light. The time independent Hamiltonian is modelled by a lattice Hamiltonian of fermions with  uniform 
nearest neighbour (NN) hoppings  given by
\begin{align}
 H^{\rm{hc}} = -t \sum_{\langle i,j\rangle,\sigma}  c_{i\sigma}^\dagger c_{j\sigma}~. \label{eq:H0}
\end{align}
The effect of the radiation can be taken into account by the vector potential ${\mathbf A}(\tau)  = A_0(\cos\omega\tau,\sin\omega\tau)$. The Hamiltonian, using Peierls substitution, is then given by
\begin{align}
H^{\rm{hc}}(\tau)= -t \sum_{\langle i,j\rangle} e^{-i\alpha\sin(\omega\tau-2\pi l/3)} c_i^\dagger c_j~.
\end{align}
where $l=0,1,2$ for the three NNs in the honeycomb lattice, i.e. $R_j=R_i+\delta_l$ and $\alpha = A_0a_0$ with $a_0$ being the lattice constant. We have  dropped the spin index as the Hamiltonian is the same for either spin sector. The Fourier components of the Hamiltonian are,
\begin{align}
 H_{n}^{\rm{hc}} = -t\sum_{\langle i,j\rangle} e^{i\frac{2\pi n l}{3}} ~J_{n}(\alpha) c_i^\dagger c_j, \label{Bessel}
\end{align}
where $J_n$ is the Bessel function of order $n$. Using Eq.~(\ref{BWexp}) one obtains the effective B-W Hamiltonian upto the first order in $(t/\omega)$ as 
\begin{align}
  H_{\rm{BW}}^{\rm{hc}} =& -\sum_{\langle i,j\rangle} J c_i^\dagger c_j  +\sum_{\langle\langle i,j\rangle\rangle} i\nu_{ij}\Lambda c^\dagger_ic_j,
\end{align}
where, 
\begin{align}
J = tJ_0(\alpha),~~\Lambda = -\frac{t^2}{\omega}\sum_{n\ne0}\frac{J_n^2(\alpha)}{n}\sin\frac{2\pi n}{3}.
\end{align}
$\nu_{ij}=\pm1$ depending on whether the next to nearest neighbour (NNN) hopping is clockwise or anticlockwise. The first term represent a renormalized hopping amplitude, whereas the second term can open a gap in the system, driving the system to the topological regime.

\section{Systems with spin-orbit coupling}\label{sec:III}
In this section, we introduce a generic 2D Hamiltonian on a  honeycomb lattice  to describe systems with spin-orbit (SO) coupling as well as to allow a buckled structure where the atoms of the sub-lattices are separated in the direction perpendicular to the plane of the lattice. Materials such as silicene, germanene and stanene can be effectively described by such a model. Cold-atom systems can  also  be used to simulate these kinds of  effective models.

The SO coupling can be introduced by adding a next to nearest neighbor (NNN) term in the Hamiltonian~\cite{Konschuh_2010}
\begin{align}
H^{\rm{SO}} =& \frac{i\lambda}{3\sqrt{3}} \sum_{\langle\langle i,j\rangle\rangle \sigma}
\sigma \nu_{ij} c_{i\sigma}^\dagger c_{j\sigma}.
\end{align}
$\lambda$ controls the strength of the SO coupling, $\sigma$ is the spin index and stands for 
$\uparrow$ and $\downarrow$ as indices and $\pm1$ in equations. We note that this takes into account only the time-reversal (TR) invariant \textit{intrinsic} SO coupling. The other prominent SO effect, Bychkov-Rashba effect, has been neglected in the following discussions and is expected 
to be small in the  systems of our interest\cite{Liu_2011}.

The staggered sub-lattice potential originating from a buckled structure can be represented as an onsite potential (taken to be uniform for simplicity)\cite{Ezawa_2011a} given by
\begin{align}
H^{\rm{ST}} =& \sum_{i\sigma}(\zeta_i lE_z - \mu) c_{i\sigma}^\dagger c_{i\sigma},
\end{align}
where $2l$ is the separation between the atoms on the 
$A$ and $B$ sub-lattices and $E_z$ is the applied electric field. $\zeta_i = +1/-1$ for $A/B$ sub lattices. The  full  Hamiltonian is thus
\begin{align}\label{hcb}
 H = H^{\rm{hc}}+H^{\rm{SO}}+H^{\rm{ST}}~.
\end{align}

We briefly note that the low energy limit of the above Hamiltonian near the $K$ and $K'$ points in the Brillouin zone has a Dirac 
 structure given by 
\begin{align}
H^\eta_\sigma &= \begin{pmatrix}\Delta^\eta_\sigma-\mu &v(\eta q_x-iq_y) \\
                      v(\eta q_x+iq_y) & -\Delta^\eta_\sigma-\mu \end{pmatrix} 
                      \label{Eqn:sile}
\end{align}
where $v=3ta_0/2$, $\Delta^\eta_\sigma =lE_z+3\sqrt{3}\eta \sigma\lambda $ and $\eta=\pm1$ are the valley indices for the two valleys $K$ and $K'$ (see Fig.~\ref{static}) at momenta $\left(\pm\frac{4\pi}{3\sqrt{3}a_0},0\right)$  . Squaring the Hamiltonian, we get the eigenvalues $E_\eta(q)=-\mu\pm\sqrt{v^2(q^2_x+q^2_y)+{\Delta^\eta_\sigma}^2 }.$ The Dirac mass term or the gap in the system is controlled by $\Delta_{\sigma}^{\eta}$. The Hamiltonian Eq.~(\ref{hcb}) is time-reversal symmetric, but the system can
be tuned  from a trivial semimetal to  a spin-hall insulating state by an  applied electric field 
 $E_z$, by tuning  $\Delta^\eta_\sigma $ through zero.

 \begin{figure}
 \begin{center}
  \includegraphics[width=0.35\textwidth]{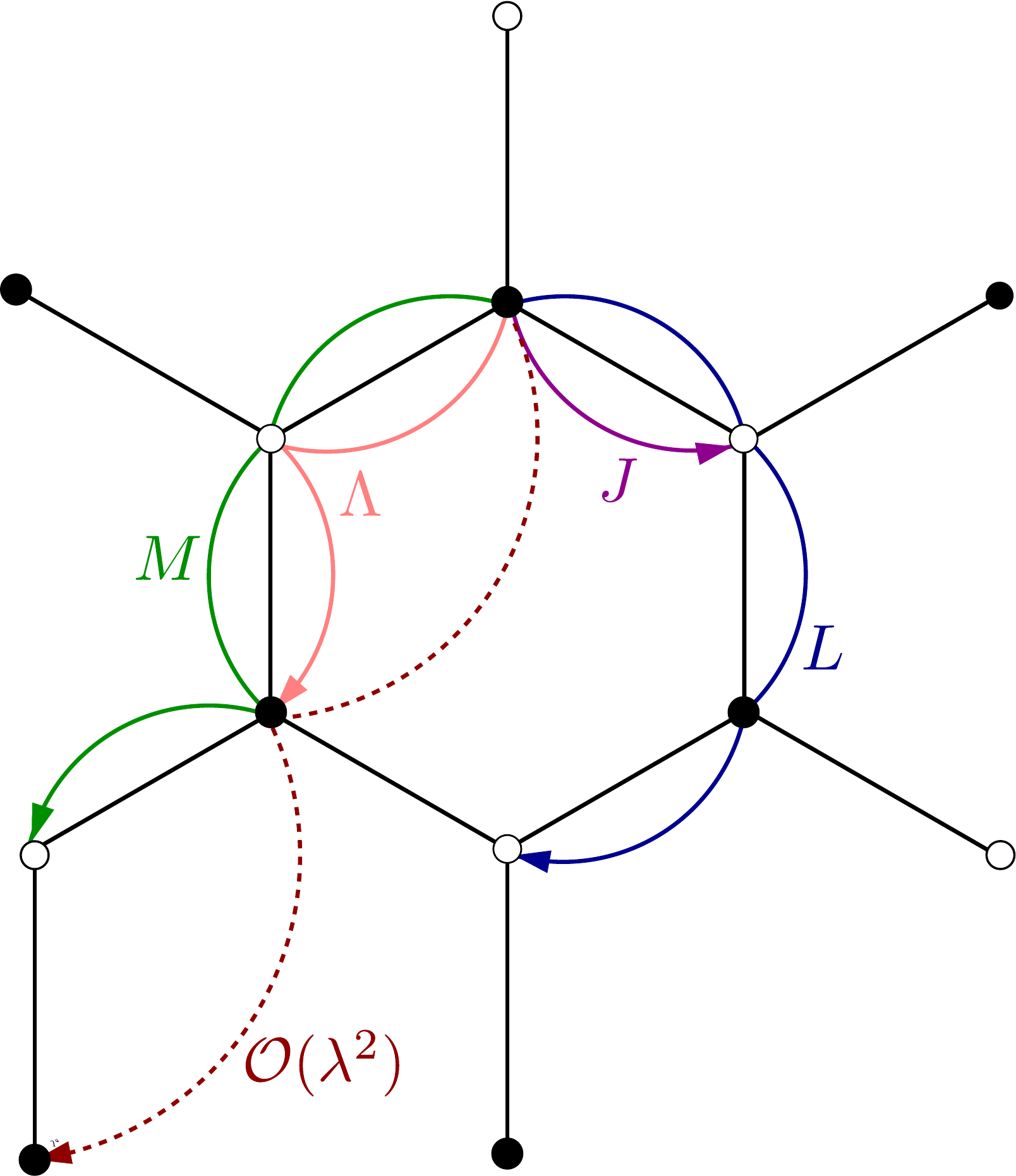}
   \end{center} 
   \caption{The various effective coupling paths in a honeycomb lattice  obtained by B-W expansion. Note that for the $L$- paths, there are two ways, both
   involving three hoppings,  to reach the $B$ sublattice from $A$. (Only one of them is shown). They contribute equally and we write them together in the amplitude Eq.~(\ref{eq:Lcoup}) that enters the Hamiltonian. The $\mathcal{O}(\lambda^2)$ contribution (dotted path) vanishes.} 
   \label{paths}
 \end{figure}

\section{B-W Expansion and Effective Hamiltonian}\label{sec:IV}
In this section we describe the procedure followed in the B-W calculation. We will start by performing a Peierls substitution on Eq.~\ref{hcb} to incorporate the effect of shining circularly polarized laser. The time dependent 
Hamiltonian thus obtained is used to calculate the Floquet Hamiltonian using Eq.~\ref{hf}. Using Eq.~\ref{BWexp}, the B-W effective Hamiltonian upto $O(\omega^{-2})$ is computed.

We first  rewrite the static Hamiltonian in Eq.~\ref{hcb} in terms of $a$ and $b$ electrons  for the $A$ and $B$ sublattice as 
\begin{align}
H \equiv&\sum_{\langle i,j\rangle \sigma} J_{\sigma} a^\dagger_{i\sigma} b_{j\sigma} +\sum_{\langle\langle i,j\rangle\rangle\sigma}( i\Lambda^0_\sigma \nu^A_{ij} +\Lambda^A)  a^\dagger_{i\sigma} a_{j\sigma} \nonumber \\
+& \sum_{i\sigma} { \tilde \mu}^A   a^\dagger_{i,\sigma} a_{i,\sigma}  + \text{all terms with}~ a,A\leftrightarrow b,B \label{Eqn:siliceneH1}
\end{align}
with $J_\sigma= - t$, $\Lambda^0_\sigma  = \frac{\sigma\lambda}{3\sqrt{3}}$,  $\Lambda^{A,B} =0$, $\nu_{ij}^A=-\nu_{ij}^B=\nu_{ij}$. The reason for the introduction
of the new notation will become clear when we start computing the corrections to the various terms using the B-W expansion. In comparison with the earlier work on the honeycomb lattice, this model has a NNN term because of the spin-orbit coupling and also a potential difference between the $A$ and $B$ sublattices due to the applied electric field $E_z$. Our aim is to see how this affects  the terms in the B-W expansion.

As mentioned earlier, the effect of shining circularly polarized light  with a  vector potential ${\bf A(\tau)}$ on  the two-dimensional honeycomb lattice  is obtained by using the Peierls substitution. Note that, ${\bf A}\cdot \delta_A = \alpha\sin(\omega\tau -2\pi l/3)$ whereas ${\bf A}\cdot \delta_B = -\alpha\sin(\omega\tau -2\pi l/3)$, for the $A$ and $B$ sublattices. The band gap at the two valleys $K$ and $K'$ can be tuned by  the applied electric field $E_z$ and also by the spin-orbit coupling term $\lambda$, whose value can be changed by the time-dependent perturbation, as we shall see below. Hence, the tunability of the band gap is highly enhanced by time-dependent perturbations.

 \begin{figure}
 \begin{center}
  \includegraphics[width=0.48\textwidth]{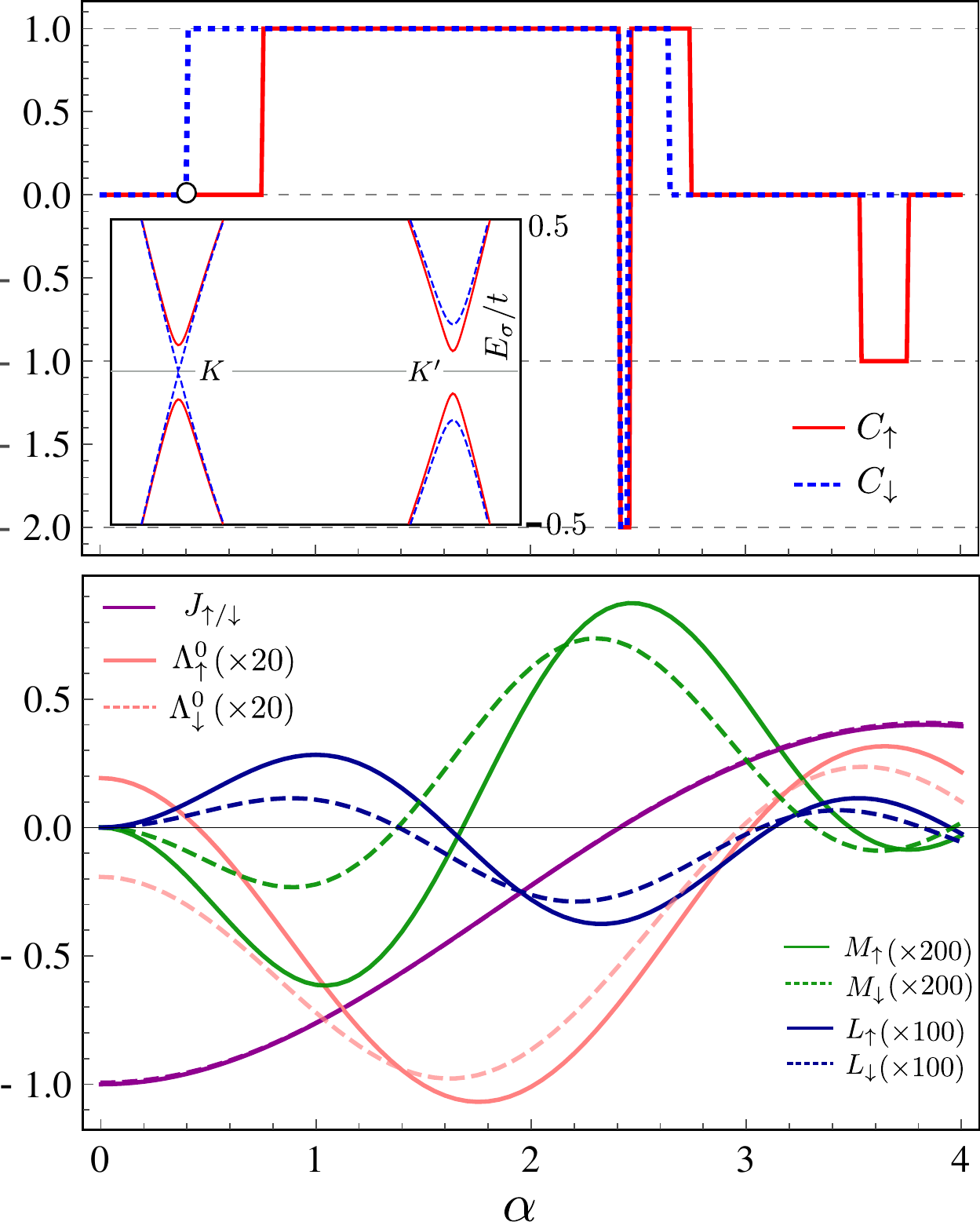}
   \end{center} 
   \caption{The Chern numbers (above) and the various amplitudes of hopping (below) are shown as a function of $\alpha=a_0A_0$. An arbitrary spin orbit coupling   $\lambda=0.05$, a staggered potential $lE_z=0.08t$ and $\omega=10$ have been used. The original hopping amplitude $t$ is taken as the unit of energy. We note that, as the hopping amplitudes differ between the two spin sectors, it is possible to achieve a spin-filtered system in the presence of SO coupling and the staggered potential. In one such situation, denoted by the white dot, we note that there is band-touching only for the $\downarrow$ spin, whereas the $\uparrow$ spin is in the gap. The band-structure near the $K,K'$ point is shown in the inset. For further  discussion of the Chern numbers, see section~\ref{sec:VI}.} 
   \label{amplitude}
 \end{figure}
 
The electric field from the irradiation couples with both the NN and the NNN hopping. Among the nearest neighbor (NN) hopping terms, the sites in the $A$ and $B$ sub-lattices have three neighbors each. In addition, the Peierls substitution has to be performed for the six next  to nearest neighbor (NNN)  sites in both the sub-lattices. The computation is more tedious than that of the NN case. The Floquet Hamiltonian is calculated by integrating the resulting 
time dependent Hamiltonian using Eq.~\ref{hf}.

 
We now use the B-W expansion defined in Eq.~(\ref{BWexp}) to obtain the effective Hamiltonian to $\mathcal{O}(\omega^{-2})$ in real space. The real space expansion is specially useful for obtaining a physical understanding of the perturbation. In realistic materials, the intrinsic spin-orbit coupling can range from a few milli-electron volts (silicene) to a few tens of milli-electron volts (germanene and stanene)~\cite{Liu_2011a,Liu_2011b}. The band-width of these materials, on the other hand, are of the order of a few electron volts. This difference in magnitude allows   us to neglect higher orders terms in $\lambda$, at higher orders in  $\mathcal{O}(1/\omega)$, while showing the results below. In general,  such approximations are not necessary,  and the B-W effective Hamiltonian can be obtained exactly at  each order, particularly for numerical purposes. We briefly sketch the  procedure
for this  in the Appendix.

With this in mind, we  compute all the terms  to $\mathcal{O}(1/\omega)$ and find
that terms of $\mathcal{O}(\lambda^2)$ cancel. To the next order, we keep only $\mathcal{O}(t^3/\omega^2)$ terms. First, the expansion renormalizes various hopping amplitudes $T_\sigma,~\Lambda^0_\sigma,~\Lambda^{A,B}_\sigma$ and ${\tilde\mu}^{A,B}_\sigma$ in Eq.~(\ref{Eqn:siliceneH1}) and we call the renormalized Hamiltonian  $H_{\rm{BW}}^I$. Second, the expansion also produces longer range hopping terms of the form:  

\noindent
\begin{align}
H_{\rm{BW}}^{II} =& \sum^{L-{\rm path}}_{i,j,\sigma} L_{\sigma} a_{i\sigma}^{\dagger} b_{j\sigma}+ \sum^{M-{\rm path}}_{i,j,\sigma} M_{\sigma} a_{i\sigma}^{\dagger} b_{j\sigma} + \rm{h.c.} \label{Eqn:siliceneH2}
\end{align}
The different $L$ and $M$ paths as well as the nearest neighbor $J$ and next nearest neighbor $\Lambda$ paths are shown in Fig.~\ref{paths}. 
The total effective B-W Hamiltonian is then
 \begin{align}
 H_{\rm{BW}} =& H_{\rm{BW}}^I +H_{\rm{BW}}^{II} \label{BWeff}.
\end{align}
Explicit forms of the various hopping amplitudes are given below:
\begin{widetext}
 \begin{align}
 J_{\sigma} =& -t J_0(\alpha)+\frac{4t\sigma \lambda }{3\omega}\sum_{n\ne0}\beta_n\sin{\frac{\pi n}{6}} +\frac{t^3}{\omega^2} \left[\sum_{n\ne0} \gamma_n\left(2\cos{\frac{2\pi n}{3}+3}\right) +\sum_{m,n \ne 0}\chi_{nm}\left( 4\cos{\frac{2\pi n}{3}+1} \right)\right], \label{Eq:J} \\
 \Lambda^0_{\sigma} =& \frac{\sigma \lambda J_0(\alpha\sqrt{3})}{3\sqrt{3}}-\sum_{n\ne0} 
 \frac{t^2J^2_n(\alpha)}{\omega n}\sin{\frac{2\pi n}{3}}
 , \\
 L_{\sigma} =&-\frac{4t\sigma\lambda}{3\omega}\sum_{n\ne0} \beta_n\sin{\frac{\pi n}{2}} +\frac{2t^3}{\omega^2} \left(\sum_{n\ne0} \gamma_n\cos{\frac{2\pi n}{3}} +\sum_{m,n \ne 0}\chi_{nm} \cos{\frac{2\pi(m-n)}{3}} \right), \label{eq:Lcoup} \\
 M_{\sigma}=& -\frac{2t\sigma\lambda}{3\omega}\sum_{n\ne0} \beta_n\cos{\pi n}\,\sin{\frac{\pi n}{6}}
 +\frac{t^3}{\omega^2} \left(\sum_{n\ne0}\gamma_n\cos{\frac{2\pi n}{3}} 
 +  \sum_{m,n \ne 0}\chi_{nm} \cos{\frac{2\pi(m+n)}{3}} \right), \\
 \Lambda^{A/B} =& -\frac{t^2 \,(\pm lE_z -\mu) }{\omega^2} \sum_{n\ne0} \frac{J^2_n(\alpha)}{n^2}\cos{\frac{2\pi n}{3}}, ~~\tilde{\mu}^{A/B} =\left(1- \frac{3t^2}{\omega^2} \sum_{n\ne0} \frac{J^2_n(A)}{n^2}\right) (\pm lE_z -\mu),
\end{align}
\end{widetext}
where $\beta_n = J_n(\alpha)J_n(\alpha\sqrt{3})/\sqrt{3}n$, $\gamma_n = J^2_n(\alpha)J_0(\alpha)/n^2$ and $\chi_{nm}=J_m(\alpha)J_n(\alpha)J_{m+n}(\alpha)/mn$.

We mention here a few important points to be noted. The presence of the SO coupling gives rise to  spin-dependent nearest neighbor hopping amplitudes $J_{\sigma}$. Furthermore, the NNNN  hopping amplitudes, the  $L$ and $M$ terms,  also become spin-dependent. The staggered onsite electric field $E_z$ plays an important role in controlling the NNN hopping amplitudes $\Lambda^{A,B}$ but appears only as a second order (in $1/\omega$) contribution. Various amplitudes have  been shown  in Fig.~\ref{amplitude}, where we note that by controlling a single parameter, $\alpha$ (which controls the strength of the driving term), their strengths can be tuned and can give rise to topological phase transitions.

Next, we proceed to write the Hamiltonian in momentum space by Fourier transforming the B-W effective Hamiltonian, Eq.~\ref{BWeff}. Alternatively, the B-W expansion can also be performed directly in the momentum space, which we have  briefly sketched in the Appendix. In the basis of the sublattices, in the spin sector $\sigma$, the B-W Hamiltonian has the form
\begin{align}
H_{\rm{BW}\sigma} &= \left(\begin{array}{cc}
                                   \delta_{\Lambda\sigma} + \xi_{A} + \tilde{\mu}^{A}& \delta_{J\sigma}+\delta_{L\sigma}+ \delta_{M\sigma}\\
                                   \delta_{J\sigma}^* +\delta_{L\sigma}^*+ \delta_{M\sigma}^* & -\delta_{\Lambda\sigma} + \xi_{B} +\tilde{\mu}^{B}
                                  \end{array}
\right), \label{eq:BWmom}
\end{align}
where
\begin{widetext}
\begin{align}
 & \delta_{J\sigma} = J_{\sigma}\left(1+2 e^{-i 3k_y a_0/2} \cos \left(\sqrt{3} k_x a_0/2\right)\right),~~ \delta_{L\sigma} =L_{\sigma} \left(e^{-3ik_y a_0} + 2\cos\left(\sqrt{3} k_x a_0\right)\right) \nonumber\\
 & \delta_{M\sigma} = 2M_{\sigma}\left(e^{-i3 k_y a_0/2} \cos \left(3 \sqrt{3}k_x a_0/2\right)+ e^{-3i k_y a_0 } \cos \left(\sqrt{3} k_x a_0\right)+e^{3 i k_y a_0/2} \cos \left(\sqrt{3} k_x a_0/2\right)\right)\nonumber\\
 &\delta_{\Lambda\sigma} = -4\Lambda^0_{\sigma}\sin \left(\sqrt{3} k_x a_0/2\right) \left(\cos \left(\sqrt{3} k_x a_0/2\right)-\cos \left(3 k_y a_0/2\right)\right)\nonumber\\
 {\rm and ~~~} &\xi_{A/B} = 2\Lambda^{A/B}\left(\cos \left(\sqrt{3} k_x a_0\right) + 2\cos \left(\sqrt{3} k_x a_0/2\right)\cos \left(3 k_y a_0/2\right)\right).\label{eq:deltas}
\end{align}
\end{widetext}
This gives the energy eigenvalues
\begin{align}
 & E_{\sigma}^{\rm BW} =\frac{\xi_{A}+\xi_{B} +\tilde{\mu}^{A}+\tilde{\mu}^{B}}{2} \pm\nonumber  \\ 
 &\sqrt{|\delta_{J\sigma}+ \delta_{L\sigma} + \delta_{M\sigma}|^2+\left(\delta_{\Lambda\sigma}+\frac{\xi_{A}-\xi_{B} +\tilde{\mu}^{A}-\tilde{\mu}^{B}}{2}\right)^2}.\nonumber
\end{align}
For an undoped system, $\mu=0$, $\left.\xi_{A}\right|_{\mu=0} = -\left.\xi_{B}\right|_{\mu=0} = \xi$ and $\left.\tilde{\mu}^A\right|_{\mu=0} = -\left.\tilde{\mu}^B\right|_{\mu=0} = \mu_0$. Using this,  the above expression reduces to 
\begin{align}
 \left.E_{\sigma}^{\rm BW}\right|_{\mu=0}=\pm\sqrt{|\delta_{J\sigma} + \delta_{L\sigma} + \delta_{M\sigma}|^2+\left(\xi+\delta_{\Lambda\sigma}+\mu_0\right)^2}.\nonumber
\end{align}
$\xi+\delta_{\Lambda\sigma}+\mu_0$ is the effective staggered potential and a finite $\mu$ simply shifts the energies.
\begin{figure}[ht]
 \begin{center}
  \includegraphics[width=0.45\textwidth]{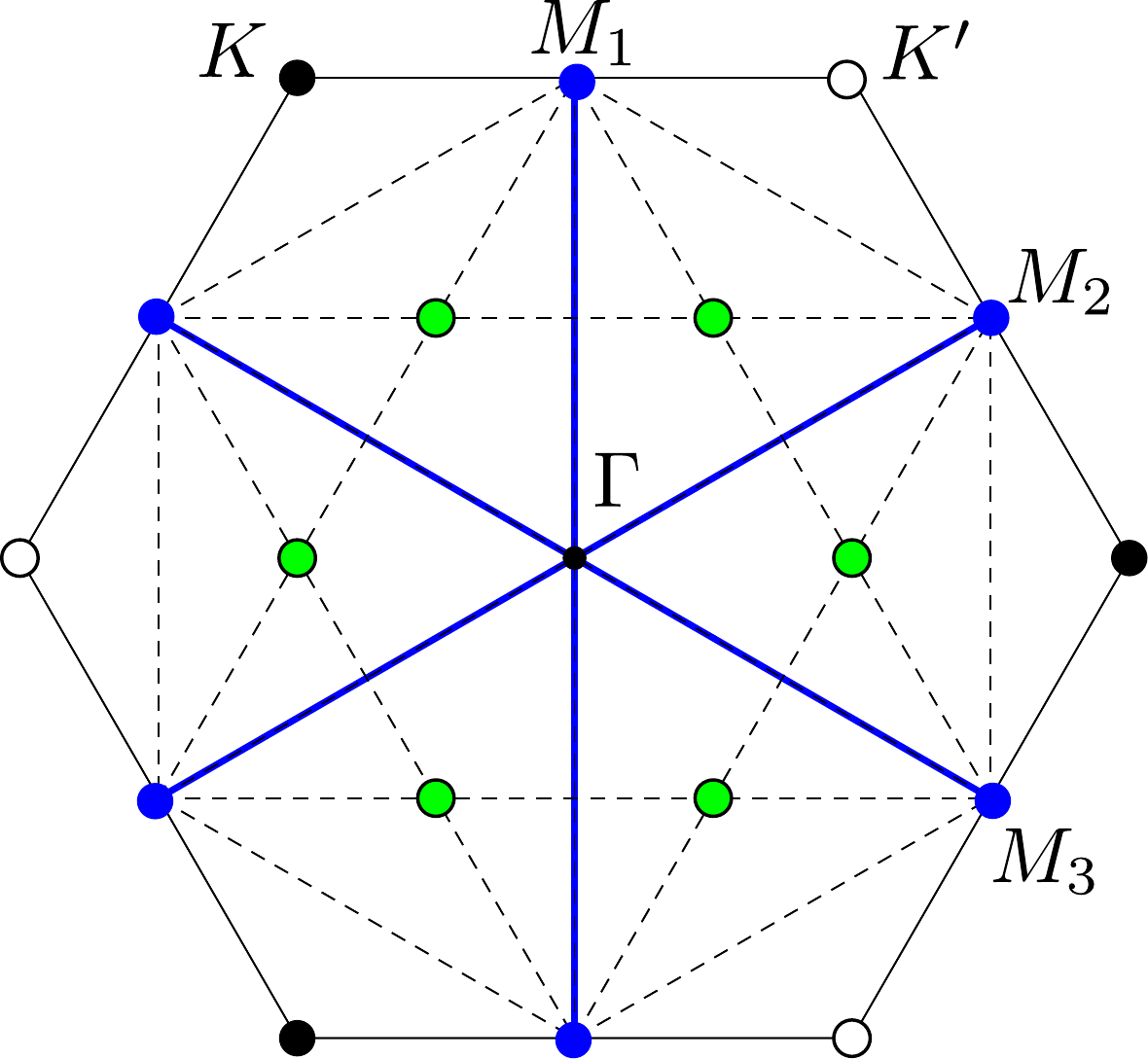}
   \caption{The various amplitudes in Eq.~(\ref{eq:deltas}) vanishes at high-symmetry points in the hexagonal Brilloiun zone. The $\delta_{\Lambda\sigma}$ term vanishes at lines (blue) joining the various $M$ points. All three $\delta_{J\sigma},\delta_{L\sigma},\delta_{M\sigma}$ terms vanish at $K$ and $K'$ points, whereas $\delta_{L\sigma}$ additionally vanishes at six other high symmetry points in the Brilloiun zone, as noted by the green dots.} 
  \label{deltas}
 \end{center} 
\end{figure}

The various amplitudes appearing in the energy expression vanishes at various high-symmetry points in the Brilloiun zone, as shown in Fig.~\ref{deltas}. As, at the $K, K'$ points,  the parameters $\delta_{J\sigma}, \delta_L, \delta_M =0$, $\delta_{\Lambda\sigma} = \pm3\sqrt{3}\Lambda_{\sigma}^0$ and $\xi=-3\Lambda^A$, the condition for a band touching point is 
\begin{align}
 &\xi+\delta_{\Lambda\sigma} +\mu_0 =0,\nonumber\\
 \Rightarrow~~~ &\mu_0 = \mp 3\sqrt{3}\Lambda_{\sigma}^0 + 3\Lambda^A.\label{eq:Kptcond}
\end{align}
Real solutions of $\omega$ from this (quadratic) equation provides the band-touching frequencies at $K/K'$ points. At the $\Gamma$ point, $\delta_{\Lambda\sigma}=0$, $\xi=6\Lambda^A$ and $\delta_{J\sigma}+ \delta_L+\delta_M =3(L_{\sigma} + 2M_{\sigma} + J_{\sigma})$. So, here the condition for band-touching is to simultaneously satisfy 
\begin{align}
 \mu_0 = -6\Lambda^A~~ {\rm and}~L_{\sigma} + 2M_{\sigma} + J_{\sigma} =0.\label{eq:Gptcond}
\end{align}
Finally, for the  various $M$ points, $\delta_{\Lambda\sigma} = 0$, $\xi = -2\Lambda^A$ and $\delta_{J\sigma}+ \delta_L+\delta_M =\pm(J_{\sigma} - 3L_{\sigma}+2M_{\sigma})$. So, for a band touching at any of the $M$ points, the condition is to simultaneously satisfy
\begin{align}\label{eq:Mpt}
 \mu_0 = 2\Lambda^A~~ {\rm and}~(J_{\sigma} - 3L_{\sigma}+2M_{\sigma})=0.
\end{align}
With appropriate limit Eq.~\ref{eq:Mpt} and Eq.~\ref{eq:Gptcond} recovers the results quoted in Ref.~\onlinecite{Mikami_2016}. A final comment is to note that as $J_{\sigma}, L_{\sigma}, M_{\sigma}, \Lambda_{\sigma}^0$ differ between the two spin sectors in the  presence of the SO coupling, it is generally not possible to have the bands touching at any of these high-symmetry points for both  the up and down spins simultaneously.

\section{The low energy limit of the B-W effective Hamiltonian }\label{sec:V}

To obtain the low energy effective Hamiltonian, we first  need to identify the band-touching points in momentum space.  In general, finding the band-touching points is not easy, because the various terms in the effective Hamiltonian are only known as a power series in the photon coupling strength. It is possible, however, to expand the Hamiltonian about a generic Dirac point, (which need not be one of the symmetric points in the Brilloiun zone) which would be useful if we could find the band-touching points. In this section we will assume that the gap closes at the $K$ and $K'$ points in the Brilloiun zone, and write down the effective Hamiltonian, so that we can compare it  with the Hamiltonian to $O(\omega^{-1})$ in the high frequency limit, obtained by 
Ezawa\cite{Ezawa_2011a}, who made this assumption.
 In the basis of the two sub-lattices,  as can be seen from Eq.~(\ref{eq:BWmom}), around the $K$ and $K'$ points, the effective Hamiltonian reduces to
\begin{align}
\left.H_{\rm{BW}}\right|_{\mathbf{k} = K/K'}&\approx \mathcal{T}_\sigma\left(\eta q_x \tau_x + q_y\tau_y\right)
 +\mathcal{D}^{\eta}_{\sigma} \tau_z -\mu \mathcal{R} I, \label{Eqn:Hbw_le}
\end{align}
with
\begin{align}
 \mathcal{T}_\sigma &= \frac{3a_0}{2}\left( 2 L_\sigma-J_\sigma +M_{\sigma}\right) \nonumber \\
 \mathcal{R} &=1+\frac{3t^2 }{\omega^2} \sum_{n\ne0} \frac{J^2_n(A)}{n^2}\left(\cos{\frac{2\pi n}{3}} -1\right) \nonumber \\
 \mathcal{D}^{\eta}_\sigma &=\left(lE_z \mathcal{R} +3\sqrt{3}\eta \Lambda^0_{\sigma} \right) \nonumber,
\end{align}
$\eta=\pm1$ for expansions around $K$ and $K'$ points respectively and $\tau_i$ are the Pauli matrices in the sub-lattice space. We note that the contributions from $L$ and $M$ paths, making the NN hopping spin-dependent was absent in Ref.~\onlinecite{Ezawa_2011a}, where the effect of the time dependent vector potential was  taken into account by Peierls substitution only in the NN hopping amplitude but not in the SO coupling. Although these contributions should be negligible in the case of silicene, it may not be small for other compounds with larger SO coupling and also for
 cold atom systems where the value of the SO coupling is arbitrary. 
We compute the eigenvalues of the Hamiltonian in Eq. (\ref{Eqn:Hbw_le}) by squaring it, and find
\begin{align}
 E_\eta(q,\sigma)&= -\mu\mathcal{R}\pm \sqrt{\mathcal{T}^2_\sigma  (q^2_x + q^2_y) +\mathcal{D}^{\eta2}_\sigma }.
 \label{Eqn:Hbw_ev2}
\end{align}
This gives the gap at the  $K/K'$ point as $2\mathcal{D}^{\pm}_{\sigma}$ for spin sector $\sigma$. The condition for the  vanishing of the gap is 
the equivalent of  the condition given in Eq.~(\ref{eq:Kptcond}) ( without taking the low energy limit). The change in sign of  the gap $\mathcal{D}^{\pm}_{\sigma}$  as  a  function of a parameter signals a  topological transition, which is characterized by the change in the spin Chern number $C_{\sigma}$ of $\pm1$.  The gap function at low energies was earlier computed by Ezawa\cite{Ezawa_2011a}. Our results agree at low values of the strength of the
electromagnetic field since the work by Ezawa\cite{Ezawa_2011a} also approximates the value of the Bessel function $J_0(A)$
 by its leading quadratic dependence on the strength of the electromagnetic field.

\begin{figure}
 \begin{center}
  \includegraphics[width=0.46\textwidth]{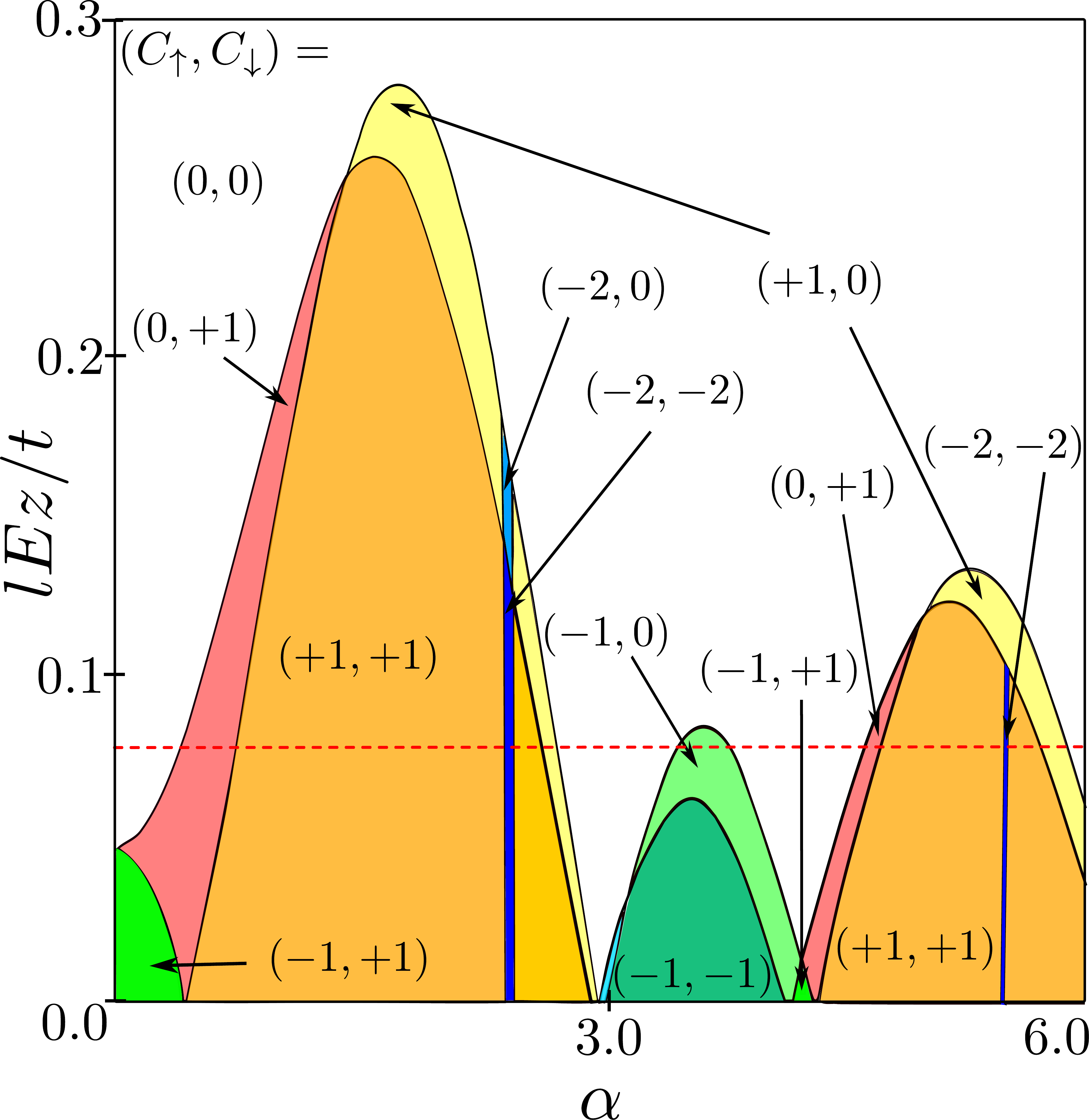}
  \caption{(color online) The phase diagram of the effective B-W Hamiltonian, Eq.~(\ref{BWeff}) characterized by the spin Chern numbers $(C_{\uparrow}, C_{\downarrow})$. Chern numbers along the dashed (red) line is shown in Fig.~\ref{amplitude}. We have taken a small but arbitrary spin orbit coupling constant $\lambda=0.05t$ and $\omega=10t$. We use the standard method for Chern number computation, c.f, Ref.~\onlinecite{Hatsugai_2005}.} 
  \label{phase}
 \end{center} 
\end{figure}

\section{Numerical Results}\label{sec:VI}
Although, in general time-periodic systems possess a  much richer topological classification than static systems~\cite{Rudner_2013}, the B-W Hamiltonian Eq.~(\ref{BWeff}) is an effective static Hamiltonian and allows us to study the model in terms of the standard topological classification of time independent systems. Neither Eq.~(\ref{hcb}) nor Eq.~(\ref{BWeff}) mixes the two spin sectors, so the spin Chern numbers $C_{\sigma}$, (independent for each spin), can classify the topology of the system. For  Eq.~(\ref{hcb}), which is valid in the absence of any time-dependent perturbation, time-reversal (TR) symmetry is intact, and we expect to have the total Chern number of the ground state $C = C_{\uparrow} + C_{\downarrow} =0$. This is not necessarily true  for the case of the B-W Hamiltonian in Eq.~(\ref{BWeff}), as the polarization of the time dependent field breaks the TR symmetry explicitly. 

\begin{figure}
 \begin{center}
  \includegraphics[width=0.48\textwidth]{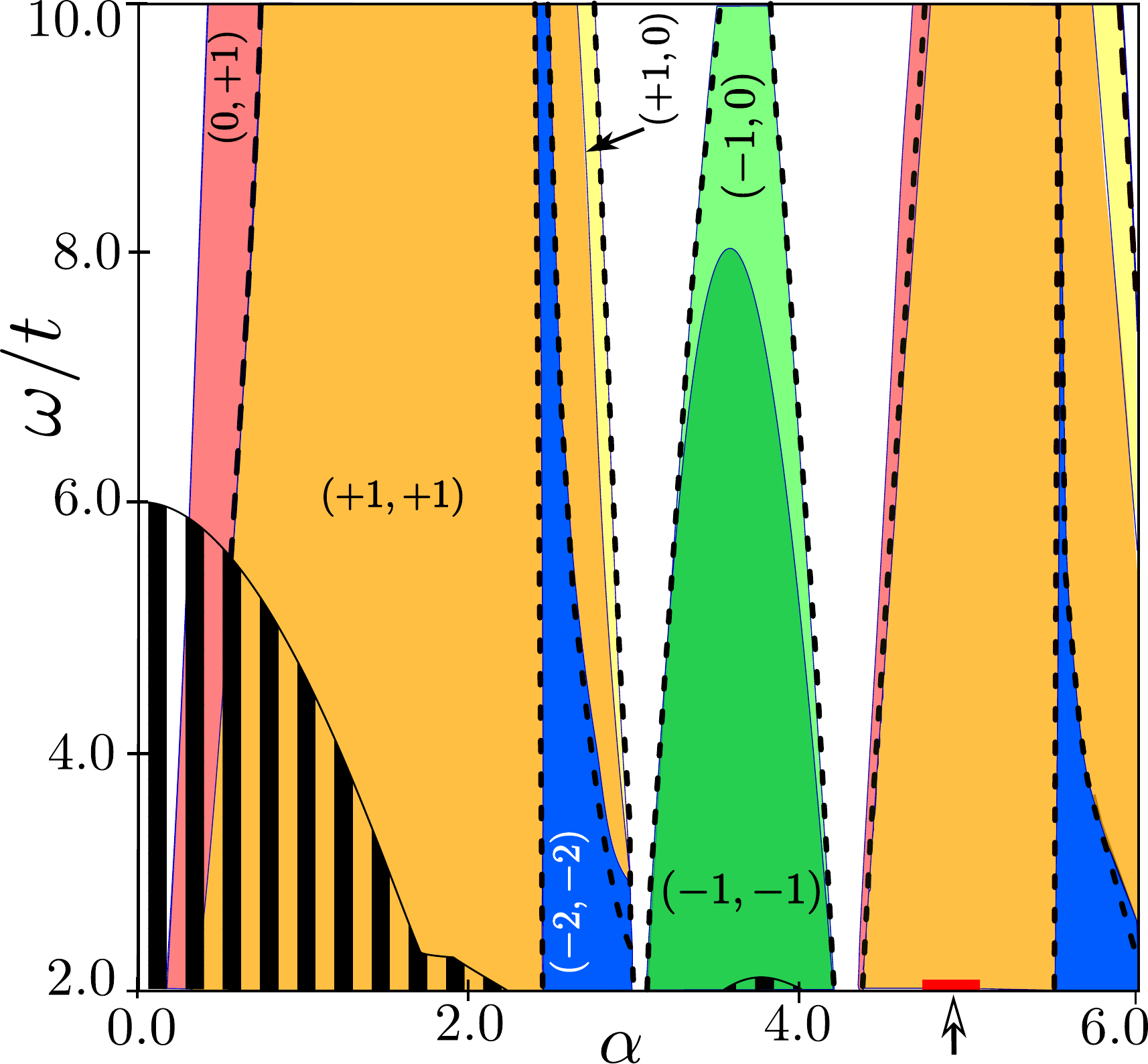}
  \caption{(color online) The phase diagram of the effective B-W Hamiltonian for both up and down spins, Eq.~(\ref{BWeff}),  with the frequency of the drive $\omega$ and the strength of the drive $\alpha$.    In the shadowed region, when the band width of the effective Hamiltonian becomes bigger than the driving frequency, the Chern number fails to match with the exact computation discussed in Sec.~\ref{sec:VI}.   We do not show other phases that appear in this shadowed region for the exact computation.
  We compare the phase boundaries of the up-spin sector of the B-W Hamiltonian with  those from the exact numerical results (indicated by dashed lines)  and see that they match exactly at high frequencies. In other regions, even for comparatively small $\omega/t$ up to 2, the match is still excellent. In the red region near $\alpha\approx 5.0$ (marked along the axis), the Chern number fails to match with exact computation, which is generally true for smaller $\omega/t$ as higher order expansion becomes necessary. Other phases are similar to that in Fig.~\ref{phase}.} 
  \label{validity}
 \end{center} 
\end{figure}

First, we compute the phase diagram of the static B-W Hamiltonian, and the results are shown in Fig~\ref{phase} and \ref{validity}. A phase diagram similar to that in  Fig~\ref{phase},  but  only for a much smaller range of parameters (both for the strength of the electromagnetic field or light  and the applied electric field $E_z$) was  obtained in Ref.~\onlinecite{Ezawa_2011a}.  The TR symmetric phase, $i.e$., when $C= C_{\uparrow}+C_{\downarrow}=0$ is present only when both the TR breaking vector potential of the drive or the staggered potentials are small. In most of the phase-space, $C_{\uparrow}=C_{\downarrow}$ instead. In relatively small regions of the phase-space,  it is possible to have $|C_{\uparrow}|\ne |C_{\uparrow}|$ and at the boundaries of these regions,  the gap closes for only one variant of the spin.
Now, if the Fermi energy is  in the gap of the other spin band, low energy excitations become completely spin-filtered. The size of such regions depend on the strength of the spin-orbit coupling.
 One such case is shown in Fig.~\ref{amplitude}.

To compare the Chern numbers obtained from the B-W expansion with the  Chern numbers of the time dependent system, one critical issue is that the occupations of the \textit{quasienergy levels} (defined below) are generally not known. Our approach is similar to that of Ref.~\onlinecite{Rudner_2013}, and we  compute  the Chern number of the quasienergy band below the  quasienergy $\epsilon=0$ which can also be defined in terms of the winding numbers of the time evolution operator above and below the band. For a time-periodic system on a lattice, the quasienergies $\epsilon_{n}(\mathbf{k})$ of band $n$ satisfy the Schr\"odinger equation for the Floquet Hamiltonian,
\begin{align}
 H_F(\mathbf{k},\tau) |u_{n} (\mathbf{k},\tau)\rangle = \epsilon_{n}(\mathbf{k})|u_{n} (\mathbf{k},\tau)\rangle,
\end{align}
where $H_F(\tau) = i\partial_{\tau} - H(\tau)$, $\mathbf{k}$ is the Bloch momentum and the \textit{Floquet states} $|u_{n} (\mathbf{k}, \tau)\rangle$ are time-periodic functions with the same period as that of $H(\tau)$. Numerically, the eigenstates of the time evolution operator  $U (T) = \mathcal{T}\exp[-i\int_0^{T}H(\tau)d\tau]$ ($\mathcal{T}$ represents time-ordered product and $T=2\pi/\omega$) provides the Floquet states  $|u_{n} (\mathbf{k}, 0)\rangle$. As these Floquet states are defined in the Brilloiun zone, one can compute (using the standard technique\cite{Hatsugai_2005}) the Chern number for each band. Finally we compare the Chern number of the up-spin sector obtained from the time-dependent Hamiltonian with that of the effective B-W Hamiltonian in Fig.~\ref{validity}, where the boundaries obtained from the time-dependent Hamiltonian have been shown by dotted lines. 
Note that the B-W results are given both for up-spin and down-spin, whereas to avoid cluttering the diagram, the exact results are given only for the up-spin sector.
Generally, in the large frequency regime, we expect to have excellent agreement as,  in fact, is  seen in the figure.

Note that  for silicene,  the spin-orbit coupling is one order of magnitude smaller than that shown in the figures, and hence the 
region of splitting between the up and down spins will be extremely narrow and not visible at the scales shown.
For germanene and stanene, the order of magnitude of the spin-orbit coupling is almost the same as that used in the figure, and so the phase diagram for both of them will be quite similar
to the one shown here.

As mentioned earlier, a time-periodic system possesses  a richer topological structure  than  its static counterpart~\cite{Rudner_2013}. Broadly speaking, the Chern number of our time-periodic system can be written as $C = C_0 - C_{\pi}$, where $C_0$ and $C_{\pi}$ are the number of chiral edge states (with the  $\pm$ signs for opposite chiralities) at the quasi-energy $\epsilon =0$ and $\omega/2$ respectively~\cite{Kundu_2014}. Starting from larger frequencies and reducing it, once the frequency becomes equal to the band-width, direct transitions from the bottom of one band to the top of the next band can occur, giving rise to band foldings. This may lead to band crossings in the extended quasi-energy zone resulting in non-zero $C_{\pi}$~\cite{Kundu_2014}. So we expect, as long as $\omega$ is larger  than the band-width, an effective Hamiltonian that is  obtained by using a  high-frequency expansion such as the B-W expansion, should reproduce the Chern number correctly. What is further interesting is that with increasing driving amplitude $A$, the electrons lose their kinetic energy ($J_{\sigma}$, Eq.~(\ref{Eq:J})), resulting in a shrinking of the band-width. This, in turn, results in a larger range of frequency where the  B-W Hamiltonian can reliably predict the Chern number.  This is shown  in Fig.~\ref{validity}, where we see that at low values
of the amplitude of the light, the  B-W expansion breaks down at $6t$, which is the band-width. But with increasing amplitude of light,
the regime of validity of the B-W Hamiltonian in Fig.~\ref{validity} increases.

Further, even if the B-W expansion does not break down at smaller frequencies, the  higher order contributions of the expansion may no longer be negligible. Such situations, where the effective Hamiltonian fails to predict the correct Chern numbers occurs with smaller values of $\omega/t$. In the Fig.~\ref{validity}, such discrepancies  occur only a very small region (red line) and extends below $\omega/t<2.$

\vspace{0.2cm}

\section{Summary and conclusion}

In summary, we have discussed a high-frequency effective Hamiltonian, using  the Brilloiun-Wigner expansion method, to describe periodically driven honeycomb lattice systems  with spin-orbit coupling and staggered potentials. Our effective Hamiltonian successfully predicts the topological nature of the system for a wide range of parameters and also provides the opportunity to explore non-trivial topological phases with external controls.

Although the B-W and other  similar high-frequency expansions provide effective time-independent Hamiltonians of the time periodic system, that does not necessarily mean that they can capture and predict correct physical properties. The time-periodic system is inherently a non-equilibrium system and in general possesses no  ground state. The lack of clarity of the occupation statistics of the electrons remain a critical issue to be resolved in such systems~\cite{Seetharam_2015,Torres_2014,Iadecola_2015}, which in turn may limit predictions of transport properties. If the driving frequency is much larger than the band width, then the energy absorption in the system is likely to  be negligible~\cite{Maricq_1982,Bukov_2015,Dehghani_2014}, and in this limit the system might be represented as  being in  quasi-equilibrium, at least for a finite time~\cite{Mori_2015}. In this case, it can be described by an effective Hamiltonian such as the B-W Hamiltonian. Nevertheless,  it may be  interesting to see  how well the transport properties as computed from a  B-W Hamiltonian compares with the other methods of computing non-equilibrium transport of the time-dependent system. We keep such studies for future.

\section*{Acknowledgments}
A.~K. was supported in part at the Technion by a fellowship of the Israel Council for Higher Education. We would also  like to thank Udit Khanna for many
useful discussions.

\appendix

\section*{Appendix: B-W Expansion in Momentum Space}
Here we briefly mention an alternative path to obtain Eq.~(\ref{eq:BWmom}) by performing the B-W expansion directly in  momentum space. First, we note that for a function $ f(t) = \exp\left(-i\gamma_1\sin \phi - i\gamma_2 \cos\phi\right),$ 
one can write,
\begin{align}
 f(t)  =& ~e^{-i\sqrt{\gamma_1^2+\gamma_2^2}\sin \left(\Omega t + \tan^{-1}\frac{\gamma_2}{\gamma_1}\right)} \nonumber \\
 =& \sum_{m=-\infty}^{\infty}J_{-m}\left(\nu\right)e^{im \left(\Omega t + \chi\right)},
\end{align}
where $\nu = \sqrt{\gamma_1^2+\gamma_2^2}$ and $\chi = \tan^{-1}\frac{\gamma_2}{\gamma_1}$. We also use $J_n(-x) = J_{-n}(x)$. Its Fourier coefficients then are
\begin{align}
 f_n =& \int_0^T\frac{dt}{T}e^{in\Omega t}f(t)\nonumber\\
 =&\sum_{m=-\infty}^{\infty}J_{-m}\left(\nu\right)e^{im  \chi}\delta_{n,-m}=~J_{n}\left(\nu\right)e^{-in \chi}.
\end{align}
Now, the irradiated silicene Hamiltonian is of the following form:
\begin{align}
 H(\mb{k},t) = \left(\begin{array}{cc}
                \xi (\mb{k},t) & \delta(\mb{k},t)\\
                \delta(\mb{k},t)^* & -\xi (\mb{k},t)
               \end{array}\right)
\end{align}
with 
\begin{widetext}
 \begin{align}
  \delta(\mb{k},t) =& t\left[e^{-i\left(\frac{\sqrt{3}}{2}k_x + \frac{3}{2}k_y\right)a_0}e^{-i\left(\frac{\sqrt{3}}{2}\cos\Omega t + \frac{1}{2}\sin\Omega t\right)\alpha} +  e^{-i\left(-\frac{\sqrt{3}}{2}k_x + \frac{3}{2}k_y\right)a_0}e^{-i\left(-\frac{\sqrt{3}}{2}\cos\Omega t + \frac{1}{2}\sin\Omega t\right)\alpha} + e^{i\alpha\sin\Omega t}\right]\nonumber\\
  \xi (\mb{k},t) =& i\Lambda\left[e^{-i\left(\frac{\sqrt{3}}{2}k_x-\frac{3}{2}k_y\right)a_0} e^{-i\left(\frac{\sqrt{3}}{2}\cos\Omega t-\frac{3}{2}\sin\Omega t\right)\alpha}- e^{i\left(\frac{\sqrt{3}}{2}k_x+\frac{3}{2}k_y\right)a_0}e^{i\left(\frac{\sqrt{3}}{2}\cos\Omega t+\frac{3}{2}\sin \Omega t\right)\alpha} - e^{-i\sqrt{3}k_xa_0}e^{-i(\sqrt{3}\cos \Omega t)\alpha}\right.\nonumber \\
  & + \left.e^{i\sqrt{3}k_xa_0}e^{i(\sqrt{3}\cos \Omega t)\alpha}+e^{-i\left(\frac{\sqrt{3}}{2}k_x+\frac{3}{2}k_y\right)a_0}e^{-i\left(\frac{\sqrt{3}}{2}\cos\Omega t+\frac{3}{2}\sin\Omega t\right)\alpha} - e^{i\left(\frac{\sqrt{3}}{2}k_x-\frac{3}{2}k_y\right)a_0}e^{i\left(\frac{\sqrt{3}}{2}\cos\Omega t-\frac{3}{2}\sin\Omega t\right)\alpha}\right]\nonumber,
 \end{align}
whose Fourier coefficients are 
\begin{align}
 &\delta_n(\mb{k}) = tJ_{n}(\alpha)\left[2e^{-i3k_ya_0}\cos\left(\frac{\sqrt{3}}{2}k_xa_0 + \frac{n\pi}{3}\right) +1 \right] ~~{\rm and} \nonumber \\
  &\xi_n(\mb{k}) = 2\Lambda J_{n}(\sqrt{3}\alpha)\left[e^{i\frac{3}{2}k_ya_0}\sin\left(\frac{\sqrt{3}}{2}k_xa_0 - \frac{n\pi}{6}\right) + e^{-i\frac{3}{2}k_ya_0}\sin\left(\frac{\sqrt{3}}{2}k_xa_0 + \frac{n\pi}{6}\right) - \sin\left(\sqrt{3}k_xa_0 + \frac{n\pi}{2}\right)\right].\nonumber
\end{align}
\end{widetext}
This defines 
\begin{align}
 H_n(\mb{k}) = \left(\begin{array}{cc}
               \xi_n (\mb{k}) & \delta_n(\mb{k})\\
               \delta_n(\mb{k})^* & -\xi_n (\mb{k})
              \end{array}\right).
\end{align}
Using $H_n$, one can then  obtain various terms of the B-W expansion using Eq.~(\ref{BWexp}).

\end{document}